\RequirePackage[2020-02-02]{latexrelease}
\documentclass[aps,showpacs,draft,twocolumn]{revtex4}  
\usepackage{graphicx}  
\usepackage{dcolumn}   
\usepackage{bm}    
\usepackage{amssymb}   
\usepackage{multirow}
\begin{document}
\widetext

\title{The effect of cutoff dependence on heavy quark spin symmetry partners}
\author{Duygu Yıldırım}
\email{yildirimyilmaz@amasya.edu.tr}%
\affiliation{Physics Department, Faculty of Sciences and Arts,  Amasya University,  Amasya, Turkey}

\date{\today}

\begin{abstract}

Hadron spectroscopy is revealed by observing heavy resonances. Among various explanations of the internal structure of these hadronic states, hadronic molecules play a unique role. For hadronic molecules, which are associated with meson-meson or meson-baryon interactions, the $\Lambda$ cutoff is a significant factor in determining the composite states' binding energies and overall properties. The cutoff becomes important when it comes to the location of hadronic molecules' masses because it influences the predictions. From this perspective, in the light of cutoff dependency,  heavy quark spin partners of near-threshold the $\chi_{c0}(3915)$,  $\chi_{c1}(3872)$,  $P_c(4440)$, and $P_c(4457)$ resonances, which are considered as hadronic molecules,  are examined.

\end{abstract}

\maketitle

In the last two decades,  a large number of experimental discoveries, especially in the heavy quarkonium sector, do not fit the expectations of the quark model~\cite{GELLMANN1964214}, which is very successful in explaining resonances so far. While those not provided by the quark model are called exotic, many theoretical models, such as the molecular picture, hadrocharmonium picture,  and diquark picture, have been proposed to describe the structure of exotic hadrons. These definitions are basically given as follows. The molecular picture proposes that exotic hadrons are composite objects made up of two or more hadrons bound together by the strong force. The hadrocharmonium picture presents that the heavy-quark pair forms a compact core around which the light quarks form a quantum-mechanical cloud. The diquark picture proposes that exotic hadrons are made up of two quarks bound together in a color anti-triplet state. These models help us understand the properties of exotic hadrons by providing a framework for interpreting experimental data on masses, production mechanisms, decay modes, and decay rates. However, no single model has yet accommodated all of the experimental data on exotic hadrons, and further research is needed to understand their properties fully. 

Hadronic molecules~\cite{Tornqvist:1993ng} attract special attention, among other definitions, because many of the seemingly unconventional resonances observed in experiments are located close to the threshold of a pair of hadrons. Therefore,  understanding the nature of hadronic molecules can help reveal the nature of the overabundant resonant structures observed worldwide in the last two decades.  There are several states that are considered as hadronic molecule candidates, including the $\chi_{c0}(3915)$~\cite{PhysRevLett.104.092001},  $\chi_{c1}(3872)$~\cite{PhysRevLett.91.262001},  $Z_c(3900)$, $Z_c(4020)$~\cite{PhysRevLett.111.242001},  $Y(4140)$~\cite{PhysRevLett.102.242002},  $P_c(4312)$~\cite{PhysRevD.100.014022}, $P_c(4440)$~\cite{PhysRevLett.124.072001}, $P_c(4457)$~\cite{PhysRevD.100.014021}, $Z_b(10610)$~\cite{Wang:2014gwa} and $Z_b(10650)$~\cite{PhysRevD.87.074006} states. Heavy meson molecules are a theoretical prediction of hadronic physics, and they are composite systems made of two or more hadrons bound together via the strong interactions. They can also be seen as analogues of light nuclei,  such as the deuteron~\cite{Tornqvist:1993ng}. 

The $\chi_{c1}(3872)$ is a strong candidate for a bound state of heavy mesons as a hadronic molecule. Recent discoveries of the $P_c(4440)$ and $P_c(4457)$ resonances also suggest that they fit the molecular description.  While the $\chi_{c0}(3915)$ and $\chi_{c1}(3872)$ are considered as $D^{(*)}\bar{D}^{(*)}$, and the $P_c(4440)$ and  $P_c(4457)$ are considered as $\bar{D}^{(*)}\Sigma^{(*)}$ because they reside so close related thresholds.  Although the $Z_b(10610)$ and $Z_b(10650)$ are highly likely that they are $B^{(*)}\bar{B}^{(*)}$ molecules,  it is unclear whether the $Z_b(10610)$ and $Z_b(10650)$ should be interpreted as a molecular state because both states have experimental results above the mass threshold of approximately $3$ MeV~\cite{Yildirim:2023znd}; therefore,  they are not examined here.  
 
Effective field theories(EFTs) are commonly used in hadron physics and provide a systematic way to describe the interactions of hadrons at low energies. The cutoff, denoted as $\Lambda$,  is an essential parameter introduced in effective field theories to handle divergences that arise during the calculation of physical quantities, particularly in loop diagrams. It serves as a momentum cutoff in loop integrals, acting as an artificial boundary that separates contributions from the low-energy (low-momentum) scale, characterizing the system from all other hadronic scales regarded as hard. A too-low $\Lambda$ value can lead to unreliable predictions due to insufficient regularization; conversely, a too-high cutoff may introduce unphysical artifacts. Guided by theoretical considerations, phenomenological constraints, and experimental data comparisons,  therefore the selection of $\Lambda$ is crucial. The choice of this cutoff is a delicate balance, as it directly impacts the reliability of predictions.  While there's no unique cutoff value, it should not surpass the breakdown scale of the EFT.  Thus,   any value is acceptable if the associated error remains within the theoretical uncertainty at the given order~\cite{Epelbaum:2006pt}.  In essence, the careful consideration and selection of the cutoff parameter are crucial for obtaining physically meaningful and reliable results in effective field theories. In this paper,  with the help of heavy quark spin symmetry it is shown that  the careful consideration and selection of the cutoff parameter are critical for getting physically significant and reliable results in heavy molecules.

The Lippmann-Schwinger equation is a fundamental tool in scattering theory, which relates the scattering amplitude to the underlying interaction potential. It provides a way to solve the Schrödinger equation for a given potential. The Lippmann-Schwinger equation is particularly useful in studying near-threshold states, where the scattering amplitude is dominated by the interaction between the two particles. In this regime, the scattering amplitude can be related to the bound state wave function through the use of the Lippmann-Schwinger equation. This allows for a detailed study of the $\chi_{c0}(3915)$,  $\chi_{c1}(3872)$,  $P_c(4440)$, and $P_c(4457)$ resonance properties of near-threshold. The equation is also used in effective field theories to relate the scattering amplitude to the underlying effective Lagrangian, which allows for a systematic expansion of the scattering amplitude in powers of the momentum.  During this study, interactions other than contact interactions are ignored because they are the next leading order~\cite{Nieves:2012tt}.

Hadronic molecules can be treated to a good approximation as composite systems, and they differ from other hadrons in that they are multi-hadron bound states that appear close to or in between two-particle thresholds. This makes possible for using nonrelativistic effective field theories to systematically access the production, decay processes, and other reactions involving hadronic molecules. At this point, the Lippmann-Schwinger equation comes to the rescue for a better understanding of heavy-bound states. To search for these bound states, we should solve the Lippmann-Schwinger equation,  which is given as, 

\begin{equation}\label{eq.1}
 T= V+V G_0 T \,,
\end{equation}
where $T$ is T-matrix,  $V$ the EFT potential and $G_0$ the resolvent operator given as $1/ (E-H_0)$.  In momentum space, the Lippmann-Schwinger equation is,

\begin{equation}\label{eq.2}
    \phi(k)+\int \frac{d^3p}{(2\pi)^3}\langle k\vert V \vert p \rangle \frac{\phi(p)}{B+\frac{p^2}{2\mu}}=0 \,,
\end{equation}
where $\phi(k)$ is the vertex function, $B$  the binding energy, and $\mu$ the reduced mass of the two hadron system. A regulator function should be used to deal with the ultraviolet divergences that arise in loop diagrams.  For the regularization of the contact range potential with $f(x)$ regulator function is in the following way,

\begin{equation}  \label{eq:3}
  \langle k\vert V \vert p \rangle = C(\Lambda) f(\frac{k}{\Lambda})f(\frac{p}{\Lambda})  \, , 
\end{equation}
where $f(x)=e^{-x^2} $ is the UV Gaussian regulator function.  The $\Lambda$ dependency of low energy constant $C(\Lambda)$ is their unwanted nature. Even if its dependency is refitted by low energy constants, the choice of cutoff can still affect the results of the calculation.   On the other hand,  in the end,  it is expected that the dependence on the cutoff should disappear in the final physical observables.  Phenomenologically, any cutoff value is acceptable if the uncertainty from the cutoff is in the order given. The energy scale of these effective field theories is typically characterized by the mass of the heavy particle involved. For charm quarks, the typical energy scale might be on the order of the charm quark mass, and for bottom quarks, it would be on the order of the bottom quark mass. In this work, the cutoff values are taken at about heavy quark mass,  that the system is included. Nevertheless, choosing any value is permitted as long as it maintains short-range effects and preserves the EFT.  Additionally, it should be mentioned that hadronic systems are states with sizes larger than the confinement scale; however, when considering interactions between hadrons at low energies, it's often helpful to describe them using EFTs, where the details of the underlying dynamics are integrated out, and encoded in a set of low-energy constants. 
 
Heavy quark spin symmetry (HQSS) is a symmetry of the strong interaction that arises in the limit of infinitely heavy quarks. In this limit, the spin of heavy quarks decouples from the system and is conserved individually. As a result, the total angular momentum of the light degrees of freedom becomes a good quantum number as well. This gives rise to the heavy quark spin symmetry~\cite{ISGUR1990527}.  It is important to note that HQSS is only approximations of symmetries. Because in the real world, quarks are not infinitely heavy.  Heavy quark effective field theory allows one to systematically include corrections that emerge from finite quark masses in a systematic expansion in $\Lambda_{QCD}/m_Q$, where $m_Q$ denotes the heavy quark mass~\cite{PhysRevD.85.114037}. Therefore, some deviation from the heavy quark limit is expected, with an error amount of approximately $\Lambda_{QCD}/m_Q$ for the LECs. As a result, a deviation of roughly $20\%$ is expected in the charm sector. 
 
HQSS implies that pseudoscalar $P$ and vector $P^*$ are degenerate and, thus, can be put into one $2\times 2$ nonrelativistic superfield matrix and expressed as~\cite{Falk:1992cx}
\begin{equation} \label{eq:4}
H=\frac{1}{\sqrt{2}}[P+\vec{P}^*\cdot \vec{\sigma}] \, ,
\end{equation}
where $\vec{\sigma}$ is the Pauli matrices. The superfield respects heavy quark rotations.  With the  $H$ superfield,  the short range interaction of the meson-meson system  without derivatives can be written as~\cite{AlFiky:2005jd}

\begin{equation}
\begin{array}{l}
\mathcal{L}_{4H} = D_{0a}Tr[\bar{H}^{(Q)} H^{(Q)} \gamma_{\mu}] Tr[H^{(\bar{Q})} H^{(\bar{Q})} \gamma^{\mu}]+ D_{0b}Tr[\bar{H}^{(Q)} H^{(Q)} \gamma_{\mu} \gamma_5] Tr[H^{(\bar{Q})} H^{(\bar{Q})} \gamma^{\mu}\gamma_5]  \\[1ex]
\hspace{0.9cm}+E_{0a}Tr[\bar{H}^{(Q)}\vec{\tau} _i H^{(Q)} \gamma_{\mu}] Tr[H^{(\bar{Q})}\vec{\tau}_i \bar{H}^{(\bar{Q})} \gamma^{\mu}] + E_{0b}Tr[\bar{H}^{(Q)} \vec{\tau}_i H^{(Q)} \gamma_{\mu} \gamma_5] Tr[H^{(\bar{Q})} \vec{\tau}_i \bar{H}^{(\bar{Q})} \gamma^{\mu}\gamma_5] \, ,    \label{5}
\end{array}
\end{equation}  
where $D_{0i}$ and $E_{0j}$ are low energy coupling constants which are unknown.  $\gamma^{\mu}$ and  $ \gamma_5$ are the Dirac gamma matrices,  $\tau_i$ are isospin matrices.  This Lagrangian gives six meson-meson molecules, and their consequences of varying cutoff values can be seen in Table \ref{tab:table1}.
 
\begin{table*}
\caption{\label{tab:table1} The HQSS-derived potential of the heavy meson system's leading order contact range relies on a linear combination of two coupling constants: $C_{0a}$ and $C_{0b}$.These coupling constants are determined from reproducing masses of the $\chi_{c0}(3915)$ and $\chi_{c1}(3872)$ for every $0.5$, $1.0$, $1.5$, $2.0$  $4.0$,  $6.0$ and $8.0$ GeV cutoff values.  Mass results are given in MeV. }
\begin{ruledtabular}
\begin{tabular}{lcccccccc|ccc}
Molecule  & $J^{P}$ & Exp.  &V &  $E_{0.5}$ & $E_{1.0}$& $E_{1.5}$& $E_{2.0}$& $E_{4.0}$ & $E_{6.0}$ &  $E_{8.0}$ & Threshold \\[1ex] \hline 
$D\bar{D}$ & $0^+$  & -  & $C_{0a}$ &  $3708$ & $3713$ & $ 3716$   & $3718$  &$3725$  & $3730$   & $3733$  &  3734 \\
$D^{*}\bar{D}$  & $1^+$ &$\chi_{c1}(3872)$  & $C_{0a}+C_{0b}$ & Input   & Input &  Input & Input  &  Input & Input  &  Input & 3876 \\[1.0ex]  
$D^{*}\bar{D}$ &  $1^+$ & - & $C_{0a}-C_{0b} $ & $3816$ & $3821$ & $ 3824$ & $3827$ & $3834$  & $3840 $  & $3846$  & 3876 	\\  
$D^{*}\bar{D}^{*}$ & $0^+$& $\chi_{c0}(3915)$ & $C_{0a}-2C_{0b}$ & Input   & Input  & Input  &Input  & Input & Input  & Input & 4017 \\[1.0ex]  
$D^{*}\bar{D}^{*}$ & $1^+$ & $X(3940)$  & $C_{0a}-C_{0b}$ & $3956$ & $3958 $ & $3959$ & $3959$ &  $3958$  & $3955$  & $3952$ & 4017	\\  
$D^{*}\bar{D}^{*}$ & $2^+$ & - &$C_{0a}+C_{0b}$  &$ 4012$  & $4012$ & $4011$ &$4019$ &	$4006$ & $4001$  &  $3995 $ &4017\\ [1.0ex] 

\end{tabular}
\end{ruledtabular}
\end{table*}

As for the heavy baryon respecting HQSS, the superfield consisting of total spin $S=\frac{1}{2}$ ground state $\Sigma$  and  $S=\frac{3}{2}$ excited state $\Sigma^*$ are given as a $2\times 3$ matrix ~\cite{CHO1994683}, 
 \begin{equation}\label{eq:6}
\vec{S}_c=\frac{1}{\sqrt{3}} \vec{\sigma}\Sigma_c +\vec{\Sigma}^*_c \, .
\end{equation}
With these superfields, the Lagrangian containing the contact range interaction without derivatives can be written as \cite{PhysRevD.98.114030}
\begin{equation} \label{eq:7}
\mathcal{L}= C_a \vec{S}^{\dagger}\cdot \vec{S} Tr[\bar{H}^{\dagger}\bar{H}]+C_b\sum_{i=1}^{3}\vec{S}^{\dagger}\cdot (J_i \vec{S}) Tr [\bar{H}^{\dagger} \sigma_i \bar{H}]  \, ,
\end{equation}
where $C_a$ and $C_b$ are low energy coupling constants. $J_i$ with $i=1,2,3$ is the spin-1 angular momentum matrices. For simplicity, isospin is ignored. This Lagrangian leads to seven contact $\bar{D}^{(*)}\Sigma^{(*)}$ molecules, and the obtained masses of them are given in Table \ref{tab:table2}.

The poles found in this study correspond to the masses of hadronic molecules. For a single channel, if the interaction is attractive and strong enough to form a bound state, the pole will be located below threshold on the first Riemann sheet.  If it is not strong enough, the pole will move onto the second Riemann sheet as a virtual state, still below threshold~\cite{Cincioglu:2016fkm}.  The corresponding contact range potentials are given in Table~\ref{tab:table1}-\ref{tab:table2}.

\begin{table*}
\caption{\label{tab:table2} The HQSS-derived potential of the heavy meson and heavy baryon system's leading order contact range relies on a linear combination of two coupling constants: $C_{a}$ and $C_{b}$.These coupling constants are determined from reproducing masses of the $P_c(4440)$ and $P_c(4457)$ for each $0.5$, $1.0$, $1.5$, $2.0$,  $4.0$,  $6.0$ and $8.0$ GeV cutoff values.  Mass results are given in MeV.}
\begin{ruledtabular}
\begin{tabular}{lcccccccc|ccc}
Molecule  & $J^{P}$ & Exp.  & V & $E_{0.5}$ & $E_{1.0}$& $E_{1.5}$& $E_{2.0}$& $E_{4.0}$ &$E_{6.0}$ & $E_{8.0}$& Threshold \\[1ex] \hline 
$\bar{D}\Sigma_c$ & $\frac{1}{2}^-$ & $ P_c(4312)$  & $C_a$ &  $4312$  &  $4314$  &  $4315$   & $4316$    &  $4319$  &  $4321$ & $ 4321$ & $4322$ \\[1ex] 
$\bar{D}\Sigma_c^*$ & $\frac{3}{2}^-$ &$ P_c(4380)$ & $C_a$   &  $ 4377$   &  $4378 $   &$4379$   & $4379$   & $4382$   &  $ 4384$  &   $4385$  & $4386$ \\[1ex] 
$\bar{D}^*\Sigma_c$ & $\frac{1}{2}^-$ &  $ P_c(4457)$   & $C_a -\frac{4}{3}C_b$  &  Input   &    Input &  Input    & Input  & Input  & Input & Input &  $4463$ \\[1ex] 
$\bar{D^*}\Sigma_c$ & $\frac{3}{2}^-$ & $ P_c(4440)$  & $C_a+\frac{2}{3}C_b$   &   Input &   Input      & Input   & Input & Input   & Input &  Input &  $4463$   \\[1.2ex] 
$\bar{D}^*\Sigma_c^*$ & $\frac{1}{2}^-$ & -& $C_a-\frac{5}{3}C_b$    &  $4501$  & $4501$    & $4501$  &$4499$   &$ 4497$  &  $4495 $  & $4493 $  &  $4527$ \\[1ex] 
$\bar{D}^*\Sigma_c^*$ & $\frac{3}{2}^-$ &-  & $C_a-\frac{2}{3}C_b$    & $4511$  & $4511$  &  $4511$& $4510$ &  $4509$  & $4507 $ &   $ 4505$   &   $4527$  \\ [1.0ex]
$\bar{D}^*\Sigma_c^*$ & $\frac{5}{2}^-$ &  -& $C_a+C_b$  & $4524$ & $4524$  &$4524$  & $4523$     & $4522$   &  $4521$ &  $ 4520$ &   $4527$   \\[1.0ex] 
\end{tabular}
\end{ruledtabular}
\end{table*}

To examine cutoff dependency on the heavy quark spin symmetry partners, masses of some observed states are taken as inputs.  Then, with the help of obtained low energy constants, in other words, binding energy,  other partners are predicted for various cutoff values.  For the charm sector,  the $\chi_{c1}(3872)$  and  $\chi_{c0}(3915)$ resonances are used as input, and their results are given in the Table~\ref{tab:table1}.  Another possible molecules are the $P_c(4440)$ and $P_c(4457)$ resonances, closely located their thresholds.  Their masses also are used as data,  the obtained results are shown in the Table~\ref{tab:table2}.  For numerical calculation,   $m_{\chi_{c1}(3872)}=3871.6$ MeV,  $m_{\chi_{c0}(3915)}=3921.7$ MeV,  $m_{P_c(4440)}=4440.3$ MeV and $m_{P_c(4457)}=4457.3$ MeV values are used~\cite{Workman:2022ynf}. To test cutoff' impact on the heavy partners,  $\Lambda$ is varied from $0.5$ to $8.0$ GeV.  The limit is cut at $8.0$ GeV because significant changes are not seen above the upper limit.

As expected, the cutoff effects are absorbed by the low energy constants. This situation might give credit for contact interactions. Because as seen Table \ref{tab:table1}-\ref{tab:table2}, as $\Lambda$ increases most states mass' increase.  It should be pointed out that they are all located under the related thresholds.   It is seen that the general behavior changes when $\Lambda$ cutoff values are taken above the threshold mass value.  Some states become more bound, and others do not.  For the $\chi_{c1}(3872)$ heavy partners, except $D^*\bar{D}^*$ molecules,  mass values increase even beyond the charm mass threshold. This behavior is expected and means that they live first Riemann sheet as a  bound state.  But for $D^*\bar{D}^*$, molecules also masses tend to decline ~\cite{Cincioglu:2016fkm}.  A similar behavior pattern is seen at the $P_c(4440)$ and  $P_c(4457)$. While the masses of  $\bar{D}^* \Sigma_c^*$ partners keep declining, other partners keep rising.

To summarize, the $\Lambda$ cutoff is critical in hadronic physics, particularly in the context of effective field theories, because it is a regularization parameter that affects the theory's predictions, including the masses and properties of hadronic molecules. It needs to be chosen carefully to ensure that the theory provides physically meaningful and reasonable results. It is important to select a cutoff that is appropriate for the system being studied. At numeric calculations, the cutoff dependency is refitted into contact terms. After that, other poles are generated. The cutoff dependence is also found not to have a significant influence on the pole positions of hadronic molecules, at least in leading order. Consequently, it can be said that the dependency is not a matter of concern as far as the cutoff values are in the limit. Further studies on the next leading order, e.g., one pion exchange, might reveal the limit and dependency of the cutoff effect.

\end{document}